\documentclass[12pt,a4paper]{article}
\usepackage{graphics,epsfig,rotating}
\newcommand{\epsfigure}[3]{
  \begin{figure}
  \epsfxsize=8.6truecm
  \epsfbox{#1}
  \caption[]{\small{#2}}
  \label{#3}
  \end{figure}
}

\begin{document}
\title{Entropy puzzle in small exploding systems}
\begin{center}
{\Large \bf Entropy puzzle in small exploding systems}
\end{center}
\begin{center}
J.P.~Bondorf$^{a}$, I.N.~Mishustin$^{a,b,c}$, G.~Neergaard$^a$
\end{center}
\begin{center}
{\it
 $^a$Niels Bohr Institute, DK-2100 Copenhagen \O, Denmark\\
 $^b$Kurchatov Institute, Russian Research Centre, 123182 Moscow, Russia\\
 $^c$Institute for Theoretical Physics, Goethe University, D-60054 
     Frankfurt am Main, Germany\\
 }
\end{center}

\normalsize
\vspace{0.3cm}

\begin{abstract}
We use a simple hard-core gas model to study the dynamics of small 
exploding systems. The system is initially prepared in a thermalized state 
in a spherical container and then allowed to expand freely into the vacuum.
We follow the expansion dynamics by recording the coordinates and velocities 
of all particles until their last collision points (freeze-out). 
We have found that the entropy per particle calculated for the ensemble of
freeze-out points is very close to the initial value. This is in apparent
contradiction with the Joule experiment in which the entropy grows when 
the gas expands irreversibly into a larger volume.
\end{abstract}

\vspace{0.5cm}

\normalsize
\baselineskip 24pt
Energetic nucleus-nucleus collisions open a unique possibility to 
study explosive dynamics of strongly interacting many-body systems in 
the laboratory. Highly excited matter produced in such collisions 
expands into vacuum until its constituents decouple (the freeze-out stage). 
There exist many models for describing this process which range 
from simple macroscopic to fully microscopic ones. 
Within thermal and fluid dynamical models it is usually assumed
that the matter expansion is isentropic, i. e. proceeds 
at constant entropy. On the other hand, as well known from statistical 
physics \cite{Lan}, only slow reversible processes conserve entropy.
It is known from the Joule experiment \cite{Jou} that the entropy
grows if the state of the system changes too fast. In this letter we 
examine the entropy conservation hypothesis on the basis of a 
microscopic model.
     
For this study we employ a simple gas model where the constituent particles 
collide like billiard balls and follow the classical Newtonian dynamics. 
This model, first introduced for simulating heavy-ion collisions in ref. 
\cite{Bon1}, was recently applied \cite{Bon2} for investigating 
de-equilibration dynamics in expanding matter. 
We consider a gas of identical balls
of radius $r_c$, which perform classical non-relativistic elastic 
scatterings at impact parameters $b<2r_c$ with conservation of energy and
momentum. Rotational degrees of freedom of the balls are ignored. 
The initial system consists of $N$ such particles placed randomly within a 
sphere of radius $R$, rejecting configurations where particles overlap within 
the hard-core distance. The particle velocities are generated from a Gaussian 
distribution with variance $T/m$ where $T$ is interpreted as temperature.
Then the particles are allowed to collide for a certain time 
(``cooking'' stage) in order to fully equilibrate the system.
For our simple interaction the total energy of the gas is obviously given 
by the ideal-gas relation $E=3NT/2$ independent of gas density.

In our simulations we arbitrarily choose conventional nuclear scales:
The mass of the constituent
particle is $m=940$ $MeV/c^2$, the hard core radius is $r_c=0.5$ fm, and the 
initial radius of the gas sphere is $R=r_0\cdot N^{1/3}$. Most simulations 
are performed with $r_0=1.2$ fm corresponding to the normal nuclear density
$\rho_0\approx 0.14$ fm$^{-3}$. 
To investigate the role played by the finite-size effects we have performed
simulations for 4 systems, N=50, 100, 200 and 400. This covers the range of 
baryon numbers actually achievable in heavy-ion collisions.
The initial average energy $E_i\approx 118$ MeV per particle and the corresponding 
temperature $T_i\approx 78$ MeV were chosen to safely ignore quantum and relativistic effects. 
The characteristic sound velocity for an ideal gas at this temparature
is $c_s\approx\sqrt{T/m} \approx 0,186c$. 

When the gas is confined in a container the particles collide not only with 
each other but also with the container wall. When the container expands the gas
particles loose energy and momentum while colliding with the moving walls.
In the case of slow expansion these losses are rapidly redistributed over all
particles and the gas remains in thermal equilibrium. This case corresponds 
to the reversible process when the temperature decreases with volume
according to the adiabatic relation
\begin{equation} \label{adia}
TV^{\gamma-1}={\rm const}~,
\end{equation}
where $\gamma\approx 5/3$ is the adiabatic index.
However, when the expansion is fast fewer particles
reach the wall and the energy losses are smaller than needed for the adiabatic
expansion. In the case of a very fast expansion of the container no particles 
can collide with the wall and therefore the energy of the gas remains constant. 
If the wall stops at a larger radius, the gas will eventually relax to a new 
equilibrium state in the larger volume. The relaxation time can be estimated 
as $\Delta R/c_s$. Since the total energy and accordingly the temperature is 
practically unchanged, the entropy of the equilibrated gas increases due to 
the larger volume, as expected for a fast irreversible process. This simple 
physics is behind the Joule experiment. Although the traditional Joule
experiment was performed with nonspherical containers, the general principles 
are obviously valid for the spherical geometry considered in this paper. 
Our simulations show that the transition from
the slow to fast expansion is rather sharp and takes place at wall 
velocities of approximately 0.5$c_s$.

More difficult problems arise when gas or fluid expand into the vacuum without 
any container. The question which we want to address is whether the 
expanding matter itself generates a sort of wall effect which may simulate 
an isentropic process. This question is closely related to the problem  
of freeze-out and collective flow in exploding systems. The simplest 
scenario which is often used in the literature is to say that the system 
expands isentropically until all interactions between the constituents 
cease. Then the change in the internal energy of the matter is transferred 
into a collective flow energy.
But the problem is that the freely expanding system has no well defined 
volume. In our previous study \cite{Bon2} we have defined the instantaneous 
volume by taking a high, 20th, moment of the particle spatial distribution.
For each time step we have defined the entropy as $S=-\sum_kp_k\ln{p_k}$ 
where $p_k$ is the occupation probability of the phase space cell $k$ in the
comoving grid. This entropy was compared with a reference entropy $S_{ref}$ 
defined for the equilibrated system of the same volume. From simulations 
at different initial conditions for a system of 50 particles we have found 
that the equilibration measure $\Sigma=\exp{(S-S_{ref})}$
at late times was not equal to 1 but rather close to 0.6.

Below we adopt a slightly different strategy using the microscopic 
information on the freeze-out field. 
The initial state is prepared in the same way as before but now after a 
``cooking'' stage the container wall is completely removed and the gas is
allowed to expand freely into the vacuum. It is important to stress that this 
free expansion starts from a state with well 
defined temperature and density. All particles are followed until their 
last collision when their coordinates and momenta are recorded. For each 
system we generate many such events and define the freeze-out field as the 
set of all such coordinates and momenta. As demonstrated in ref. \cite{Bon2}, 
these fields are nonlocal in space and time, in contrast to a simplified 
Cooper-Frye picture \cite{CoFr} assuming a sharp freeze-out hypersurface. 
We point out also that the number of 
freeze-out points per event is generally less than the number of particles 
because some particles leave the system without any collision (see Table 1 below).

After obtaining the freeze-out field we calculate the average
characteristics of the phase space occupation. Utilising spherical symmetry 
of the system we divide it into a number of spherical shells of radii $R_k$.
For each shell we calculate the average density of freeze-out points
$\rho(r)$, collective velocity $u(r)$ and temperature $T(r)$. The collective 
velocity is defined simply as the mean radial velocity of frozen-out particles
in a given shell i. e. between $R_k$ and $R_{k+1}$,
\begin{equation}
u(r)\equiv \overline{v}(r)=\frac{1}{N_k}\sum_{i=1}^{N_k} v_r({\bf r}_i)~,
\end{equation}
where $N_i$ is the number of freeze-out points in this shell, $\sum_k N_k=N$. 
The temperature is determined from the variance of velocities in the shell,
assuming the ideal-gas relation,
\begin{equation}
T(r)=\frac{m}{3}\left(\overline{v^2}(r)-\overline{v}^2(r)\right),
\end{equation}
and the mean-square velocity is defined in the standard way,
\begin{equation}
\overline{v^2}(r)=\frac{1}{N_k}\sum_{i=1}^{N_k} v^2({\bf r}_i)~.
\end{equation}

With this information in hand we can calculate the final entropy of the gas.
Since at this late stage of expansion the gas is very dilute one can use 
the ideal-gas formulae. The freeze-out entropy in a given shell is defined as 
\begin{equation} \label{ent}
S(r)=N_k\ln{\left[\frac{V_ke^{5/2}}{N_k\lambda_T^3}\right]}~,~~~ 
\lambda_T=\left(\frac{2\pi\hbar^2}{mT(r)}\right)^{1/2}~,
\end{equation} 
and the total entropy is obviously given by the sum over the shells.
We believe that this definition of entropy is valid despite the fact
that it is applied not to the real gas but to the ensemble of freeze-out 
points in the phase space. Here one can use an analogy with the microwave
background radiation in the Universe which keeps its entropy constant
despite the fact that the photons have decoupled from the matter at the
recombination stage a long time ago.

To make statistical errors similar for different systems, 
the number of generated events is chosen to be inversely proportional to the 
system's particle number (see Table 1). This guarantees that 
the total number of freeze-out points is approximately the same for all 
considered systems.  
The dynamical simulations
were performed with the time step of 0.5 fm/c which was sufficient to 
resolve practically all collisions.   

The results of the simulations are presented in Figs.~1-4 and Table 1 (see below). 
Because of the limited statistics the spatial distributions shown in the 
figures are sensitive to the binning of data. Most calculations were done
by sampling freeze-out points in spherical shells of equal volume 
(r$^3$ binning). This guarantees uniform statistical errors for 
the bulk parts of distributions but leads to enhanced fluctuations on
their tails. Moreover, for unambiguous separation of flow and thermal 
components the radial bin size should be sufficiently small. 

It is necessary to emphasise that the spatial characteristics presented 
in Figs.~1-4 correspond to sampling of freeze-out points irrespective 
of the times when particles have actually decoupled from the system. 
They represent the whole freeze-out history and in
this respect differ from the time evolution of the gas
characteristics usually presented in gasdynamical calculations.
We believe that this representation is more adequate for calculating 
observable characteristics of small exploding systems. This is especially 
true for the interpretation of experimental data on energetic 
nucleus-nucleus collisions.  

Fig. 1 shows the spatial density of freeze-out points averaged over all 
events. In all cases it has 
a bulk part and a tail. The bulk density is about 0.1 particles/fm$^{3}$ and 
almost independent of the system. This should be compared with an initial
density of 0.14 particles/fm$^3$. In the tail region the density rapidly
drops to zero over a radial distance of about 2-3 fm. Such behaviour should 
be anticipated from the general consideration of the freeze-out process 
\cite{CoFr}. The tail is formed by particles emitted through the surface 
at early times and the inner part contains particles from the bulk 
freeze-out.      
\epsfigure{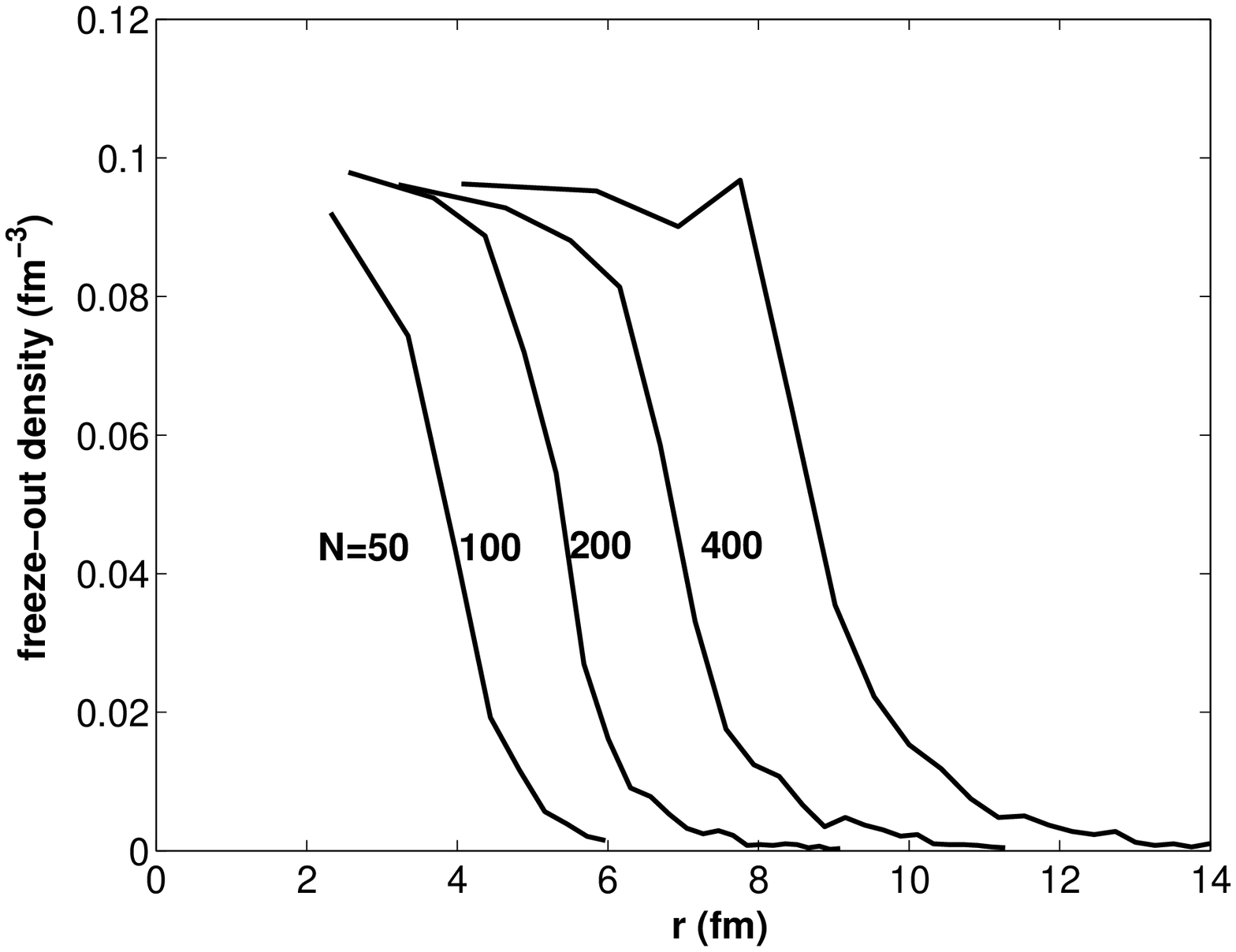}{Radial density of freeze-out points for free expansion of 
the gas spheres with $N=$50, 100, 200 and 400. The initial temperature 
78 MeV and density 0.14 fm$^{-3}$ of the gas are the same in all cases.}{figure1}

Fig. 2 presents the collective velocity profiles calculated on the freeze-out 
fields. With a certain degree of imagination one can recognize a Hubble-like
behaviour. As expected the collective velocity grows to the outer edge of 
the distribution. The peak value of about 0.5-0.6 is reached somewhere in the 
tail region. This value is in good agreement with gasdynamical 
calculations \cite{Zel} predicting for leading particles a velocity of 
about 3$c_s$.
\epsfigure{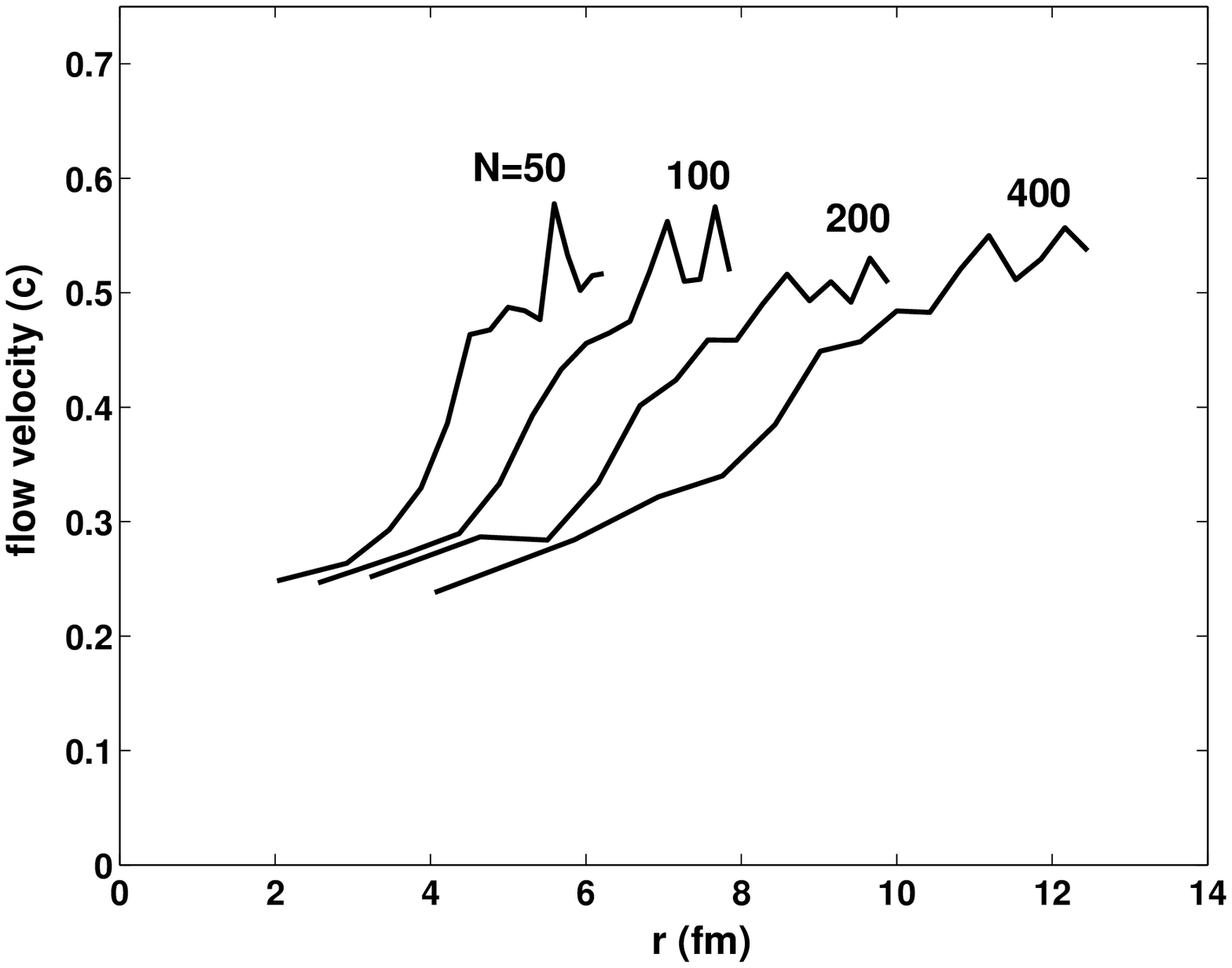}{Radial profiles of the collective velocity at freeze out. 
Notations are the same as in Fig.~1.}{figure2}

The temperature profiles presented in Fig.~3 are in a certain
sense complementary to the flow profiles. 
One can see that the temperature reaches
maximum values at the edge of the bulk region and these values decrease 
progressively with the system's size. This can be explained by the fact 
that the freeze-out process in larger systems develops at later stages 
of expansion leading to lower freeze-out temperatures. As well known (see
e.g. ref. \cite{Zel}), in a macroscopic system the freeze-out temperature 
approaches zero and the whole thermal energy is finally transformed into
collective flow. We clearly see the transition from ``small'' to ``large'' 
systems by analyzing the average number of collisions per particle. As we
see from our simulations, this number scales roughly as $\sqrt{N}$. While it is 
only about 1 for $N=50$ (small system), it is already 2-3 for $N=400$ 
(mesoscopic system), and will be about 10 for $N=5000$ (large system).  
\epsfigure{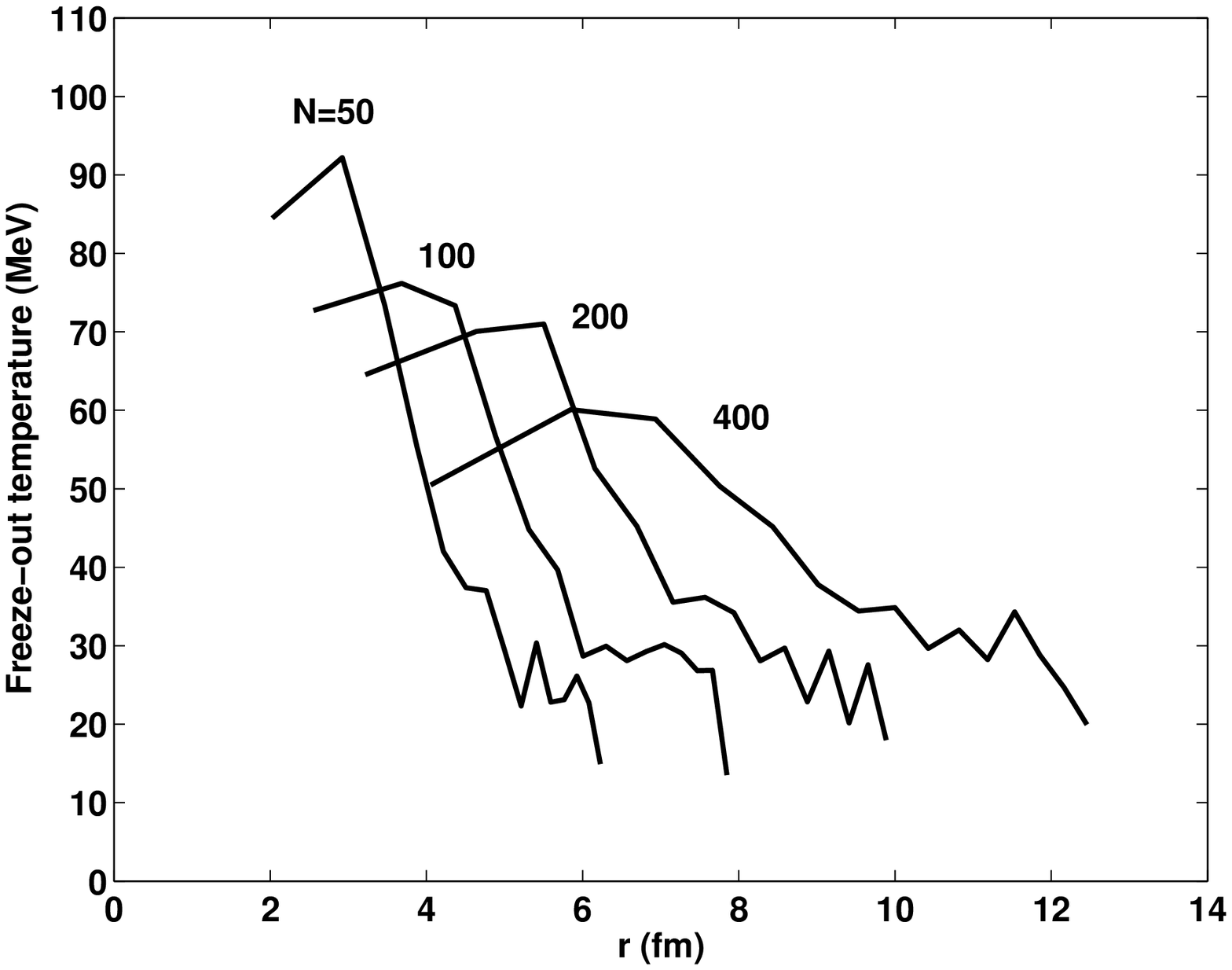}{Radial profiles of the temperature at freeze-out. Notations
are the same as in Fig.~1.}{figure3}

Finally we come to the most interesting quantity, {\it i.e.}~the entropy per 
particle as defined by eq. (\ref{ent}). Fig.~4 shows the corresponding 
profiles. One can see two clear features. First, the entropy
per particle is rather constant over the bulk region and its value varies 
very little with the  system size. Second, there is a significant rise in 
the entropy per particle in the tail region well above the bulk value 2-3.
Moreover, the smaller the system the stronger the rise. This latter 
trend can be explained by the bigger volume per particle in the outer 
tail region. 
 
Now we can go back to our discussion of entropy conservation. For this 
analysis we use the total entropy per particle calculated at freeze-out and 
compare it with the initial entropy. The latter is calculated by applying 
the same eq.~(\ref{ent}) for the whole gas in the initial volume. The results 
are presented in Table~1 together with the number of particles, the number of events,
the total number of freeze-out points for each size ($N$) and the freeze-out fraction
$F_{freeze}=n_{freeze}/(N \cdot n_{event})$. 

\begin{table}
\centering
\vspace{1cm}
\begin{tabular}{r|rrrrr}
$N$ & $n_{event}$ & $n_{freeze}$ & $F_{freeze}$ & $s_{initial}$ & $s_{final}$  \\ \hline 
50 & 128 & 3848 & 0.60 & 2.68 & 3.27 \\
100 & 64 & 4388 & 0.69 & 2.69 & 3.07 \\
200 & 32 & 4673 & 0.73 & 2.71 & 2.97 \\
400 & 16 & 5048 & 0.79 & 2.70 & 2.75 \\
\end{tabular}
\caption{\small{In this table, $n_{event}$ is the number of events,
$n_{freeze}$ is the number of freeze-out points,
$F_{freeze}$ is the freeze-out fraction defined in the text,
$s_{inital}$ and $s_{final}$ are the inital and final entropy per particle
respectively. Notice that the freeze-out fraction increases with system size,
and that the final freeze-out entropy per particle 
approaches the inital value for larger systems.}}
\end{table}

From Table 1 one can see that surprisingly enough the initial and final 
entropies per particle are rather close to each other in all cases. The 
increase of 
about 0.5 units is largest for the smallest system considered ($N=50$). 
Formally this is a 20\% effect which is quite significant. However, one 
should bear in mind that in classical statistics the absolute value of 
entropy is defined up to a constant. 
This increase in entropy for small systems is an effect of the surface of the system: 
As seen from Fig.~4, the entropy per particle increases significantly with 
radius, and since there is relatively more surface in a small system 
also the total entropy per particle is bigger.
\epsfigure{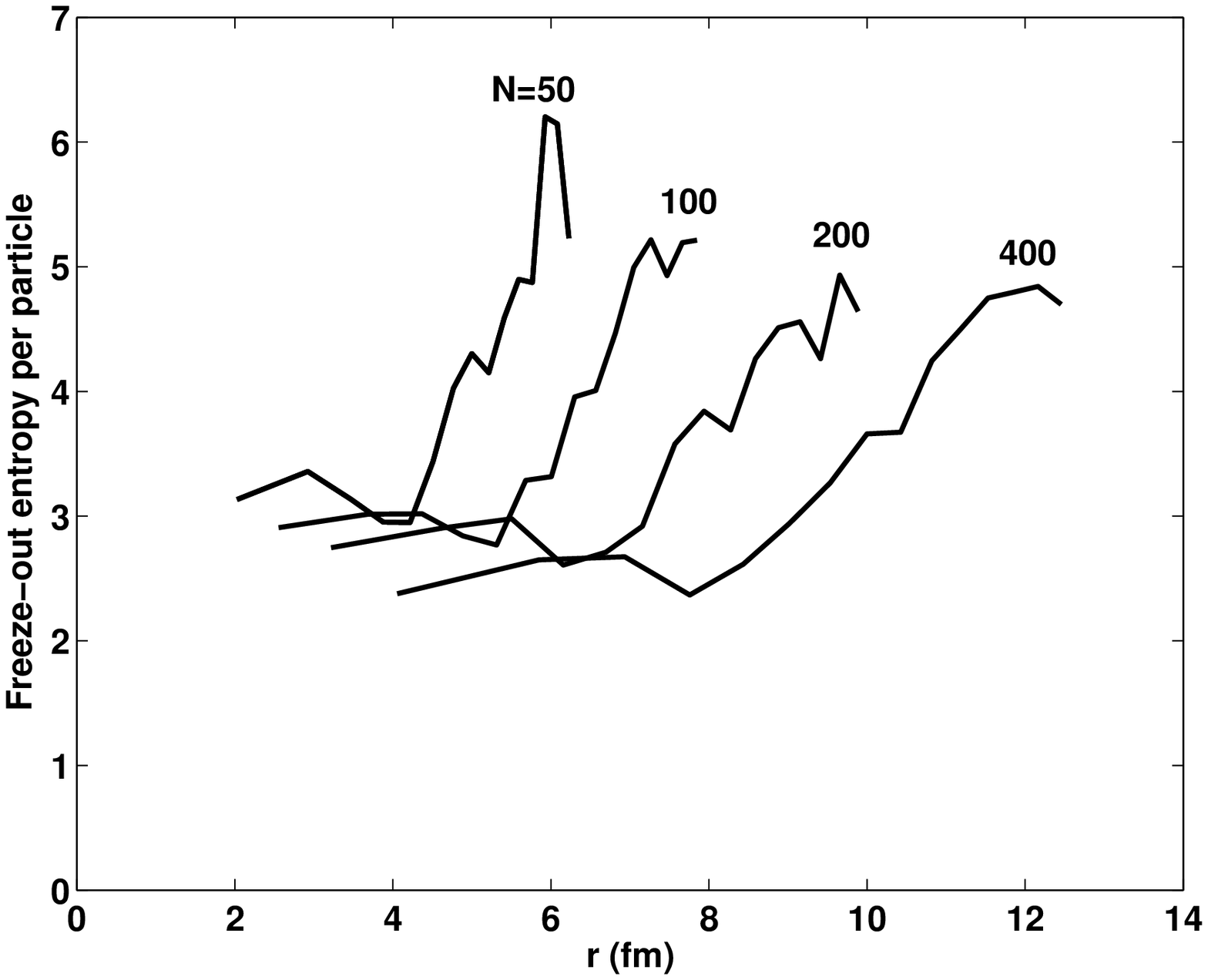}{Radial distribution of the entropy per particle at 
freeze-out. Notations are the same as in Fig.~1.}{figure4}

One may wonder, if entropy per particle is conserved, what happens to the 
total entropy of the system? Indeed, a significant fraction of particles 
$(1-F_{freeze})$ leaves the system without scatterings.  
The simulations show that these "missing" particles are emitted early in the
expansion and come predominantly from the surface region. We have not analyzed
their characteristics in detail but we know that these particles were initially
in thermal equilibrium with the rest of the system. Thus they should carry
away approximnately the same amount entropy per particle as in the initial 
state. Therefore, we expect that the total entropy of the system is also 
approximately conserved.  

In conclusion, we have used
a simple model for a repulsive gas to study the explosive
dynamics of small systems. We have demonstrated that in the course of free 
expansion the temperature drops and collective flow develops in the gas.
In contrast to our expectations 
we have found that the entropy per
particle defined on the freeze-out field is almost conserved even in systems 
with a few hundred particles. This justifies the application of thermal 
and hydrodynamic models for describing matter evolution in energetic collisions
of medium and heavy nuclei. Based on these results we put forward
a new interpretation of the old Joule experiment. The gas expansion 
in this case is approximately isentropic at freeze-out before the 
particles hit the wall. Then the entropy is produced while the system 
equilibrates in the larger volume.  

The authors thank H. Feldmeier and T, Dossing for productive discussions.
This work was supported in part by the Danish National Research Council,
Duetsche Forschung Gemeinschaft, DFG grant 436 RUS 113/711/0-1, and Russian 
Fund of Fundamental Research, RFFI grant 03-02-04007.


\begin{thebibliography}{99}

\bibitem{Lan} L.D. Landau and E.M. Lifshitz, Statistical Physics I,
Pergamon, New York, 1968.

\bibitem{Jou}  J. P. Joule, Phil. Mag. {26} 369-83 (1845) or 
Joule Scientific Papers Vol 1 172-89, Taylor and Francis, London 1884. 

\bibitem{Bon1} J.P. Bondorf, H.T. Feldmeier, S. Garpman and E.C. Halbert, 
Phys. Lett. {\bf 65B}, 217 (1976).

\bibitem{Bon2} J.P. Bondorf, H. Feldmeier, I.N. Mishustin and G. Neegaard, 
Phys. Rev. C{\bf 65}, 017601 (2002).

\bibitem{Zel} Ya.B. Zel'dovich and Yu.P. Raizer, Physics of shock waves and 
high-temperature hydrodynamic phenomena, Academic Press, New York and London, 1967.
 
\bibitem{CoFr} F. Cooper and G. Frye, Phys. Rev. D {\bf 10}, 186 (1974). 

\end{thebibliography}
\end{document}